\documentstyle[12pt,aas2pp4]{article}

\received{}
\accepted{}
\journalid{}{}
\articleid{}{}
\lefthead{Robertson S.L.}
\righthead{}

\begin{document}
\title{Bigger Bursts From Merging Neutron Stars}
 
\author{STANLEY L. ROBERTSON}

\affil{Department of Physics, Southwestern Oklahoma State University,
Weatherford, OK 73096}

\begin{abstract}
GRB 990123 may have radiated more than one solar mass equivalent in just its
gamma emissions. Though this may be within the upper limit of the
binding energies available from neutron stars in the Schwarzschild metric, it
is difficult to imagine a process with the required efficiency of
conversion to gamma rays. Neutron stars of $\sim 10 M_\odot$ are permitted
in the Yilmaz metric. A merger of two neutron stars of maximum mass could
release approximately $10 M_\odot$ equivalent binding energy.
\end{abstract}

\keywords{Gamma Ray Bursts, Black Hole Physics, Stars:
neutron}

\section{Introduction}
GRB 990123 was remarkable in several respects. A recently predicted
optical flash (Sari \& Piran 1999) was observed for it (Odewahn,
Bloom \& Kulkarni 1999). It has been associated
with a galaxy at a redshift of z=1.6 (Bloom et al. 1999).
If gamma emissions from GRB 990123 were isotropic, $\sim 3x10^{54}$ erg was
emitted just in gamma rays (Blandford \& Helfand 1999), making
it the most energetic burst observed to date. Whether or not the bursts are
beamed is an open question. At present there is no evidence of beaming
(Tavani 1998). No distribution differences
between primary bursts and afterglows (Grindlay 1999)
have yet been found. Future optical and radio observations should
soon settle this question decisively.

The optical flash prediction was based on a relativistic fireball model.
One possible generator of fireballs would be mergers of neutron stars (NSs).
Such mergers should occur at approximately the observed rate of bursts in
star forming epochs ($z \sim 1 - 3$). Unfortunately, GRB 990123
presents an energy crisis (Tavani 1998). It is unlikely that low-mass
NSs can produce the required energy. According to General Relativity,
an object of nuclear density and more than 2.8 M$_\odot$ would
be a black hole (Kalogera \& Baym 1996, Friedman \& Ipser 1987).
Although a merger of two NSs of $\sim 1.7 M_\odot$ might release
$\sim 10^{54}$ erg (see below), getting it all in gamma photons seems
very unlikely. The energy crisis can be resolved and the neutron
star merger model retained by removing the event horizon obstacle of
General Relativity.

\section{Yilmaz Neutron Stars}
Event horizons in the Schwarzschild metric (SM) of
General Relativity are indicated by the vanishing of the metric coefficient
\begin{equation}
g(r)~=~1~-~2 u(r).
\end{equation}
where
\begin{equation}
u(r)~=~\frac{GM}{c^2r}
\end{equation}
is the gravitational potential at distance r from mass M.
In addition to an isotropic static limit metric, the Yilmaz metric (YM)
(Yilmaz 1958, 1971, 1975, 1992, 1994, 1995) has a corresponding metric
coefficient of
\begin{equation}
g(r)~=~\exp(-2 u(r)).
\end{equation}
This coefficient can be derived from the principle of
equivalence and special relativity applied to frames co-moving
with an accelerated particle (Einstein 1907, Rindler 1969, Yilmaz 1975).
With metric coefficients differing only in second and higher order terms,
SM and YM differ by only a few percent to
radii as small as the innermost marginally stable orbit (Robertson
1999). This will permit decisive tests of disk accretion theories
with and without event horizons at the core. The Yilmaz theory passes
the four classic weak-field tests. It permits local energy-momentum
conservation, has no adjustable parameters, no singularities, no
event horizons and can be reconciled with quantum theory
(Alley 1995, Yilmaz 1994, 1995). Gravitationally compact
objects can exist in the YM but they are not black holes. Radially directed
photons can always escape.

Robertson (1999) has calculated maximum
NS masses for the two metrics using a simpleminded model of a non-rotating
star of constant proper density. For nuclear saturation density,
$2.8 x 10^{14}$ g/cm$^3$, a maximum NS mass of $2.4 M_\odot$
is obtained for the SM. The maximum mass found for a non-rotating
Yilmaz star is about $9 M_\odot$. Maximally rotating stars
could be about 25\% more massive (Friedman \& Ipser 1987).

If one of a pair of merging Yilmaz NSs were near maximum mass,
the binding energy released in reaching a stable state would be close
to the mass equivalent of the smaller star. Without exploring the
endless possibilities for mergers, consider the collision of
two stars of equal mass. For such pairs, Figure 1 shows the results for the two
metrics. The merger of two $9 M_\odot$ Yilmaz stars would yield $11 M_\odot$ or
$2x10^{55}$ erg; perhaps enough to relieve both the energy and
efficiency crises of the merged NS models. Athough binding energies
of the innermost marginally stable orbit differ by only 3.5\%, low mass NSs
are clearly more tightly bound in SM.

\begin{figure}
\plotone{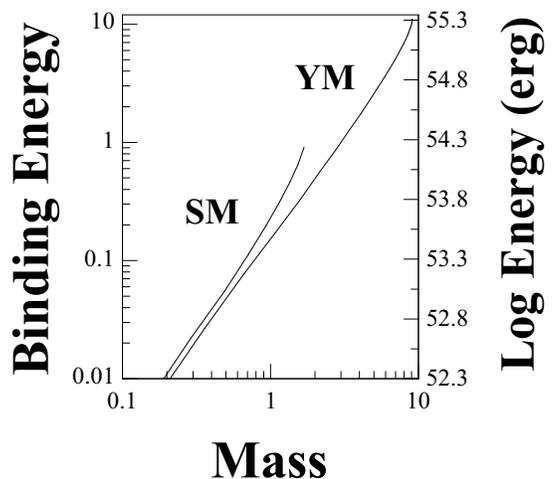}
\figcaption{ Binding energy in $M_\odot$ released for stability of merged
equal mass neutron stars as a function of the mass ($M_\odot$) of one
star. The scale at the right shows binding energy in erg.}
\end{figure}

\section{Discussion}
In addition to relieving the crises imposed by General Relativity,
the Yilmaz star model is consistent with the known properties of the
galactic black hole candidates (GBHCs). Although there are
some spectral differences between GBHCs and low mass NSs, none
have been convincingly attributed to the presence of an
event horizon. Robertson (1999) has argued that GBHCs
are simply massive NSs. Those in low mass
x-ray binary systems typically spin at a few tens of Hz and have
magnetic fields of $\sim 10^{10 - 11}$G; sufficient to suppress
surface bursts. The distribution of dynamically determined GBHC masses seems
to be limited to something less than $\sim 12 M_\odot$ with a peak near
$7 M_\odot$, (Chen, Livio \& Schrader 1997) in accord with the Yilmaz NS
model. GBHCs of $\sim 7 M_\odot$ appear to constitute a large
fraction the compact objects in low mass x-ray binaries
(e.g. Barret, McClintock \& Grindlay 1996). This suggests that a sufficient
population of such objects exists to provide blasts such as GRB 990123.

One indication that GBHCs are simply massive NSs consists of the similar
spectra of low hard states of GBHCs and Atoll class NSs.
In the low state, a magnetic propeller effect (Campana et al. 1998,
Zhang et al. 1998) disrupts the inner accretion disk, cuts
off the flow to the NS surface and generates a hard x-ray spectrum.
A disrupted inner disk has been attributed to GBHCs as well
(Done \& $\dot{Z}$ycki 1999).
The magnetic pressure at the co-rotation radius is tens of
megabar for Atolls and hundreds of megabar for Z class NSs.
For the spins and magnetic moments attributed to the GBHCs by
Robertson, many of the GBHCs would have pressures similar to the Atolls at their
co-rotation radii, thereby producing some of their similar
spectral and timing characteristics.
One would expect spectral differences for low-state black holes
which would lack such impediments to plasma flows.
Since compact objects in YM and SM differ essentially only by the presence
or absence of surfaces and magnetic fields, the YM provides a
tool capable of confirming or rejecting the existence of black holes.
If it survives the tests and resolves the GRB energy crisis, we may also
need to re-examine the luminosity-redshift relation for accelerations
of the cosmic expansion rate.

\section{Acknowledgements}
I thank Dr. Darryl Leiter for suggesting that others might be
interested in the implications of Yilmaz theory for GRBs.

\end{document}